# Data Sharing Options for Scientific Workflows on Amazon EC2

Gideon Juve, Ewa Deelman, Karan Vahi, Gaurang Mehta
*USC Information Sciences Institute*
*{gideon,deelman,vahi,gmehta}@isi.edu*

Benjamin P. Berman
*USC Epigenome Center*
*bberman@usc.edu*

Bruce Berriman
*NASA Exoplanet Science Institute, Infrared Processing and Analysis Center, Caltech*
*gbb@ipac.caltech.edu*

Phil Maechling
*Southern California Earthquake Center*
*maechlin@usc.edu*

***Abstract*—Efficient data management is a key component in achieving good performance for scientific workflows in distributed environments. Workflow applications typically communicate data between tasks using files. When tasks are distributed, these files are either transferred from one computational node to another, or accessed through a shared storage system. In grids and clusters, workflow data is often stored on network and parallel file systems. In this paper we investigate some of the ways in which data can be managed for workflows in the cloud. We ran experiments using three typical workflow applications on Amazon's EC2. We discuss the various storage and file systems we used, describe the issues and problems we encountered deploying them on EC2, and analyze the resulting performance and cost of the workflows.**

## I. Introduction

Scientists are using workflow applications in many different scientific domains to orchestrate complex simulations, and data analyses. Traditionally, large-scale workflows have been run on academic HPC systems such as clusters and grids. With the recent development and interest in cloud computing platforms many scientists would like to evaluate the use of clouds for their workflow applications. Clouds give workflow developers several advantages over traditional HPC systems, such as root access to the operating system and control over the entire software environment, reproducibility of results through the use of VM images to store computational environments, and on-demand provisioning capabilities.

One important question when evaluating the effectiveness of cloud platforms for workflows is: How can workflows share data in the cloud? Workflows are loosely-coupled parallel applications that consist of a set of computational tasks linked via data- and control-flow dependencies. Unlike tightly-coupled applications, such as MPI jobs, in which tasks communicate directly via the network, workflow tasks typically communicate through the use of files. Each task in a workflow produces one or more output files that become input files to other tasks. When tasks are run on different computational nodes, these files are either stored in a shared file system, or transferred from one node to the next by the workflow management system.

Running a workflow in the cloud involves creating an environment in which tasks have access to the input files they require. There are many existing storage systems that can be deployed in the cloud. These include various network and parallel file systems, object-based storage systems, and databases. One of the advantages of cloud computing and virtualization is that the user has control over what software is deployed, and how it is configured. However, this flexibility also imposes a burden on the user to determine what system software is appropriate for their application. The goal of this paper is to explore the various options for sharing data in the cloud for workflow applications, and to evaluate the effectiveness of various solutions.

The contributions of this paper are:

- A description of an approach that sets up a computational environment in the cloud to support the execution of scientific workflow applications.
- An overview of the issues related to workflow storage in the cloud and a discussion of the current storage options for workflows in the cloud.
- A comparison of the performance (runtime) of three real workflow applications using five different storage systems on Amazon EC2.
- An analysis of the cost of running workflows with different storage systems on Amazon EC2.

Our results show that the cloud offers a convenient and flexible platform for deploying workflows with various storage systems. We find that there are many options available for workflow storage in the cloud, and that the performance of storage systems such as GlusterFS [11] is quite good. We also find that the cost of running workflows on EC2 is not prohibitive for the applications we tested, however the cost increases significantly when multiple virtual instances are used. At the same time we did not observe a corresponding increase in performance.

The rest of the paper is organized as follows: Section II describes the set of workflow applications we chose for our experiments. Section III gives an overview of the execution environment we set up for the experiments on Amazon EC2. Section IV provides a discussion and overview of storage systems (including various file systems) that are used to communicate data between workflow tasks. Sections V and VI provide results of our experiments in terms of both runtime and cost. Sections VII and VIII describe related work and conclude the paper.

## II. Workflow Applications

In order to evaluate the cost and performance of data sharing options for scientific workflows in the cloud we considered three different workflow applications: an astronomy application (Montage), a seismology application (Broadband), and a bioinformatics application (Epigenome). These three applications were chosen because they cover a wide range of application domains and a wide range of resource requirements. Table I shows the relative resource usage of these applications in three different categories: I/O, memory, and CPU. The resource usage of these applications was determined using a workflow profiler[1], which measures the I/O, CPU usage, and peak memory by tracing all the tasks in the workflow using ptrace [27].

**TABLE I**
Application resource usage comparison

| Application | I/O | Memory | CPU |
|---|---|---|---|
| Montage | High | Low | Low |
| Broadband | Medium | High | Medium |
| Epigenome | Low | Medium | High |

The first application, Montage [17], creates science-grade astronomical image mosaics using data collected from telescopes. The size of a Montage workflow depends upon the area of the sky (in square degrees) covered by the output mosaic. In our experiments we configured Montage workflows to generate an 8-degree square mosaic. The resulting workflow contains 10,429 tasks, reads 4.2 GB of input data, and produces 7.9 GB of output data (excluding temporary data). We consider Montage to be I/O-bound because it spends more than 95% of its time waiting on I/O operations.

The second application, Broadband [29], generates and compares seismograms from several high- and low-frequency earthquake simulation codes. Each Broadband workflow generates seismograms for several sources (scenario earthquakes) and sites (geographic locations). For each (source, site) combination the workflow runs several high- and low-frequency earthquake simulations and computes intensity measures of the resulting seismograms. In our experiments we used 6 sources and 8 sites to generate a workflow containing 768 tasks that reads 6 GB of input data and writes 303 MB of output data. We consider Broadband to be memory-limited because more than 75% of its runtime is consumed by tasks requiring more than 1 GB of physical memory.

The third and final application, Epigenome [30], maps short DNA segments collected using high-throughput gene sequencing machines to a previously constructed reference genome using the MAQ software [19]. The workflow splits several input segment files into small chunks, reformats and converts the chunks, maps the chunks to the reference genome, merges the mapped sequences into a single output map, and computes the sequence density for each location of interest in the reference genome. The workflow used in our experiments maps human DNA sequences from chromosome 21. The workflow contains 529 tasks, reads 1.9 GB of input data, and produces 300 MB of output data. We consider Epigenome to be CPU-bound because it spends 99% of its runtime in the CPU and only 1% on I/O and other activities.

[1] http://pegasus.isi.edu/wfprof

## III. Execution Environment

In this section we describe the experimental setup that was used in our experiments. We ran experiments on Amazon's EC2 infrastructure as a service (IaaS) cloud [1]. EC2 was chosen because it is currently the most popular, feature-rich, and stable commercial cloud available.

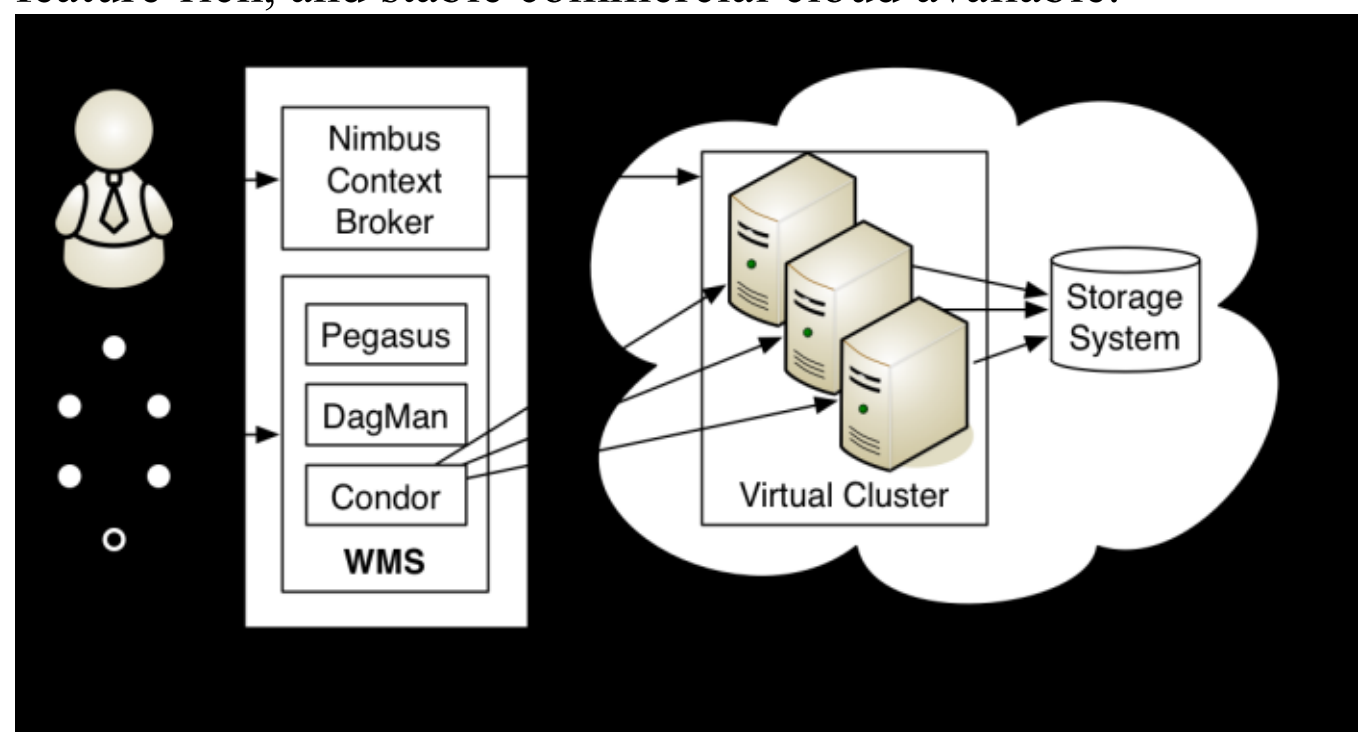

Fig. 1. Execution environment

There are many ways to configure an execution environment for workflow applications in the cloud. The environment can be deployed entirely in the cloud, or parts of it can reside outside the cloud. For this paper we have chosen the latter approach, mirroring the configuration used for workflows on the grid. In our configuration, shown in Fig. 1, we have a submit host that runs outside the cloud to manage the workflows and set up the cloud environment, several worker nodes that run inside the cloud to execute tasks, and a storage system that also runs inside the cloud to store workflow inputs and outputs.

### *A. Software*

The execution environment is based on the idea of a *virtual cluster* [4,10]. A virtual cluster is a collection of virtual machines that have been configured to act like a traditional HPC cluster. Typically this involves installing and configuring job management software, such as a batch scheduler, and a shared storage system, such as a network file system. The challenge in provisioning a virtual cluster in the cloud is collecting the information required to configure the cluster software, and then generating configuration files and starting services. Instead of performing these tasks manually, which can be tedious and error-prone, we have used the Nimbus Context Broker [18] to provision and configure virtual clusters for this paper.

All workflows were planned and executed using the Pegasus Workflow Management System [7], which includes the Pegasus mapper, DAGMan [5] and the Condor *schedd* [21]. Pegasus is used to transform a resource-independent, abstract workflow description into a concrete plan, which is then executed using DAGMan. The latter manages dependencies between executable tasks, and Condor schedd manages individual task execution. The Pegasus mapper, DAGMan, the Condor manager, and the Nimbus Context Broker service were all installed on the submit host.

To deploy software on the virtual cluster we developed a virtual machine image based on the stock Fedora 8 image provided by Amazon. To the stock image we added the Pegasus worker node tools, Globus clients, Condor worker daemons, and all other packages required to compile and run the tasks of the selected workflows, including the application binaries. We also installed the Nimbus Context Broker agent to manage the configuration of the virtual machines, and wrote shell scripts to generate configuration files and start the required services. Finally, we installed and configured the software necessary to run the storage systems that will be described in Section IV. The resulting image was used to deploy worker nodes on EC2. With the exception of Pegasus, which needed to be enhanced to support Amazon S3 (see section IV.A) the workflow management system did not require modifications to run on EC2.

### *B. Resources*

Amazon EC2 offers several different resource configurations for virtual machine instances. Each instance type is configured with a specific amount of memory, CPUs, and local storage. Rather than experimenting with all the various instance types, for this paper only the c1.xlarge instance type is used. This type is equipped with two quad core 2.33-2.66 GHz Xeon processors (8 cores total), 7 GB RAM, and 1690 GB local disk storage. In our previous work we found that the c1.xlarge type delivers the best overall performance for the applications considered here [16]. A different choice for worker nodes would result in different performance and cost metrics. An exhaustive survey of all the possible combinations is beyond the scope of this paper.

### *C. Storage*

To run workflows we need to allocate storage for 1) application executables, 2) input data, and 3) intermediate and output data. In a typical workflow application executables are pre-installed on the execution site, input data is copied from an archive to the execution site, and output data is copied from the execution site to an archive. Since the focus of this paper is on the storage systems we did not perform or measure data transfers to/from the cloud. Instead, executables were included in the virtual machine images, input data was pre-staged to the virtual cluster, and output data was not transferred back to the submit host. For a more detailed examination of the performance and cost of workflow transfers to/from the cloud see our previous work [16].

Each of the c1.xlarge instances used for our experiments has 4 “ephemeral” disks. These disks are virtual block-based storage devices that provide access to physical storage on local disk drives. Ephemeral disks appear as devices to the virtual machine and can be formatted and accessed as if they were physical devices. They can be used to store data for the lifetime of the virtual machine, but are wiped clean when the virtual machine is terminated. As such they cannot be used for long-term storage.

Ephemeral disks have a severe first write penalty that should be considered when deploying an application on EC2. One would expect that ephemeral disks should deliver performance close to that of the underlying physical disks, most likely around 100 MB/s, however, the observed performance is only about 20 MB/s for the first write. Subsequent writes to the same location deliver the expected performance. This appears to be the result of the virtualization technology used to expose the drives to the virtual machine. This problem has not been observed with standard Xen virtual block devices outside of EC2, which suggests that Amazon is using a custom disk virtualization solution, perhaps for security reasons. Amazon’s suggestion for mitigating the first-write penalty is for users to initialize ephemeral disks by filling them with zeros before using them for application data. However, initialization is not feasible for many applications because it takes too much time. Initializing enough storage for a Montage workflow (50 GB), for example, would take almost as long (42 minutes) as running the workflow using an uninitialized disk. If the instance using the disk is going to be provisioned for only one workflow, then initialization does not make economic sense.

For the experiments described in this paper we have not initialized the ephemeral disks. In order to get the best performance without initialization we used software RAID [20]. We combined the 4 ephemeral drives on each c1.xlarge instance into a single RAID 0 partition. This configuration results in first writes of 80-100 MB/s, and subsequent writes around 350-400 MB/s. Reads peak at around 110 MB/s from a single ephemeral disk and around 310 MB/s from a 4-disk RAID array. The RAID 0 disks were used as local storage for the systems described in the next section.

## IV. Storage Options

In this section we describe the storage services we used for our experiments and any special configuration or handling that was required to get them to work with our workflow management system. We tried to select a number of different systems that span a wide range of storage options. Given the large number of network storage systems available it is not possible for us to examine them all. In addition, it is not possible to run some file systems on EC2 because Amazon does not allow kernel modifications (Amazon does allow modules, but many file systems require source code patches as well). This is the case for Lustre [24] and Ceph [33], for example. Also, in order to work with our workflow tasks (as they are provided by the domain scientists), the file system either needs to be POSIX-compliant (i.e. we must be able to mount it and it must support standard semantics), or additional tools need to be used to copy files to/from the local file system, which can result in reduced performance.

It is important to note that our goal with this work is not to evaluate the raw performance of these storage systems in the cloud, but rather to examine application performance in the context of scientific workflows. We are interested in exploring various options for sharing data in the cloud for workflow applications and in determining, in general, how the performance and cost of a workflow is affected by the choice of storage system. Where possible we have attempted to tune each storage system to deliver the best performance, but we have no way of knowing what combination of parameter values will give the best results for all applications without an exhaustive search. Instead, for each storage system we ran some simple benchmarks to verify that the storage system functions correctly and to determine if there are any obvious parameters that should be changed. We do not claim that the configurations we have used are the best of all possible configurations for our applications, but rather represent a typical setup.

In addition to the systems described below we ran a few experiments using XtreemFS [14], a file system designed for wide-area networks. However, the workflows performed far worse on XtreemFS than the other systems tested, taking more than twice as long as they did on the storage systems reported here before they were terminated without completing. As a result, we did not perform the full range of experiments with XtreemFS.

### *A. Amazon S3*

Amazon S3 [2] is a distributed, object-based storage system. It stores un-typed binary objects (e.g. files) up to 5 GB in size. It is accessed through a web service that supports both SOAP and a REST-like protocol. Objects in S3 are stored in directory-like structures called *buckets*. Each bucket is owned by a single user and must have a globally unique name. Objects within a bucket are named by *keys*. The key namespace is flat, but path-like keys are allowed (e.g. "a/b/c" is a valid key).

Because S3 does not have a POSIX interface, in order to use it, we needed to make some modifications to the workflow management system. The primary change was adding support for an S3 client, which copies input files from S3 to the local file system before a job starts, and copies output files from the local file system back to S3 after the job completes. The workflow management system was modified to wrap each job with the necessary GET and PUT operations.

Transferring data for each job individually increases the amount of data that must be moved and, as a result, has the potential to reduce the performance of the workflow. Using S3 each file must be written twice when it is generated (program to disk, disk to S3) and read twice each time it is used (S3 to disk, disk to program). In comparison, network file systems enable the file to be written once, and read once each time it is used. In addition, network file systems support partial reads of input files and fine-grained overlapping of computation and communication. In order to reduce the number of transfers required when using S3 we implemented a simple whole-file caching mechanism. Caching is possible because all the workflow applications used in our experiments obey a strict write-once file access pattern where no files are ever opened for updates. Our simple caching scheme ensures that each file is transferred from S3 to a given node only once, and saves output files generated on a node so that they can be reused as input for future jobs that may run on the node.

The scheduler that was used to execute workflow jobs does not consider data locality or parent-child affinity when scheduling jobs, and does not have access to information about the contents of each node's cache. Because of this, if a file is cached on one node, a job that accesses the file could end up being scheduled on a different node. A more data-aware scheduler could potentially improve workflow performance by increasing cache hits and further reducing transfers.

### *B. NFS*

NFS [28] is perhaps the most commonly used network file system. Unlike the other storage systems used, NFS is a centralized system with one node that acts as the file server for a group of machines. This puts it at a distinct disadvantage in terms of scalability compared with the other storage systems. For the workflow experiments we provisioned a dedicated node in EC2 to host the NFS file system. Based on our benchmarks the m1.xlarge instance type provides the best NFS performance of all the resource types available on EC2. We attribute this to the fact that m1.xlarge has a comparatively large amount of memory (16GB), which facilitates good cache performance. We configured NFS clients to use the *async* option, which allows calls to NFS to return before the data has been flushed to disk, and we disabled *atime* updates.

### *C. GlusterFS*

GlusterFS [11] is a distributed file system that supports many different configurations. It has a modular architecture based on components called *translators* that can be composed to create novel file system configurations. All translators support a common API and can be stacked on top of each other in layers. The translator at each layer can decide to service the call, or pass it to a lower-level translator. This modular design enables translators to be composed into many unique configurations. The available translators include: a server translator, a client translator, a storage translator, and several performance translators for caching, threading, pre-fetching, etc. As a result of these translators there are many ways to deploy a GlusterFS file system. We used two configurations: *NUFA* (non-uniform file access) and *distribute*. In both configurations nodes act as both clients and servers. Each node exports a local volume and merges it with the local volumes of all other nodes. In the NUFA configuration all writes to new files are performed on the local disk, while reads and writes to existing files are either performed across the network or locally depending on where the file was created. Because files in the workflows we tested are never updated, the NUFA configuration results in all writes being directed to the local disk. In the distribute configuration GlusterFS uses hashing to distribute files among nodes. This configuration results in a more uniform distribution of reads and writes across the virtual cluster compared to the NUFA configuration.

### *D. PVFS*

PVFS [3] is a parallel file system for Linux clusters. It distributes file data via striping across a number of I/O nodes. In our configuration we used the same set of nodes for both I/O and computation. In other words, each node was configured as both a client and a server. In addition, we configured PVFS to distribute metadata across all nodes instead of having a central metadata server.

Although the latest version of PVFS was 2.8.2 at the time our experiments were conducted, we were not able to run any of the 2.8 series releases on EC2 reliably without crashes or loss of data. Instead, we used an older version, 2.6.3, and applied a patch for the Linux kernel used on EC2 (2.6.21). This version ran without crashing, but does not include some of the changes made in later releases to improve support and performance for small files.

## V. Performance Comparison

In this section we compare the performance of the selected storage options for workflows on Amazon EC2. The critical performance metric we are concerned with is the total runtime of the workflow (also known as the makespan). The runtime of a workflow is defined as the total amount of wall clock time from the moment the first workflow task is submitted until the last task completes. The runtimes reported in the following sections do not include the time required to boot and configure the VM, which typically averages between 70 and 90 seconds [15], nor do they include the time required to transfer input and output data. Because the sizes of input files are constant, and the resources are all provisioned at the same time, the file transfer and provisioning overheads are assumed to be independent of the storage system chosen.

In discussing the results for various storage systems it is useful to consider the I/O workload generated by the applications tested. Each application generates a large number (thousands) of relatively small files (on the order of 1 MB to 10 MB). The write pattern is sequential and strictly write-once (no file is updated after it has been created). The read pattern is primarily sequential, with a few tasks performing random accesses. Because many workflow jobs run concurrently, many files will be accessed at the same time. Some files are read concurrently, but no file is ever read and written at the same time. These characteristics will help to explain the observed performance differences between the storage systems in the following sections.

Note that the GlusterFS and PVFS configurations used require at least two nodes to construct a valid file system, so results with one worker are reported only for S3 and NFS. In addition to the storage systems described in section 4, we have also included performance results for experiments run on a single node with 8 cores using the local disk. Performance using the local disk is shown as a single point in the graphs.

### A. Montage

The performance results for Montage are shown in Fig. 2. The characteristic of Montage that seems to have the most significant impact on its performance is the large number (~29,000) of relatively small (a few MB) files it accesses. GlusterFS seems to handle this workload well, with both the NUFA and distribute modes producing significantly better performance than the other storage systems. NFS does relatively well for Montage, beating even the local disk in the single node case. This may be because we used the async option with NFS, which results in better NFS write performance than a local disk when the remote host has a large amount of memory in which to buffer writes, or because using NFS results in less disk contention. The relatively poor performance of S3 and PVFS may be a result of Montage accessing a large number of small files. As we indicated in Section IV, the version of PVFS we have used does not contain the small file optimizations added in later releases. S3 performs worse than the other systems on small files because of the relatively large overhead of fetching and storing files in S3. In addition, the Montage workflow does not contain much file reuse, which makes the S3 client cache less effective.

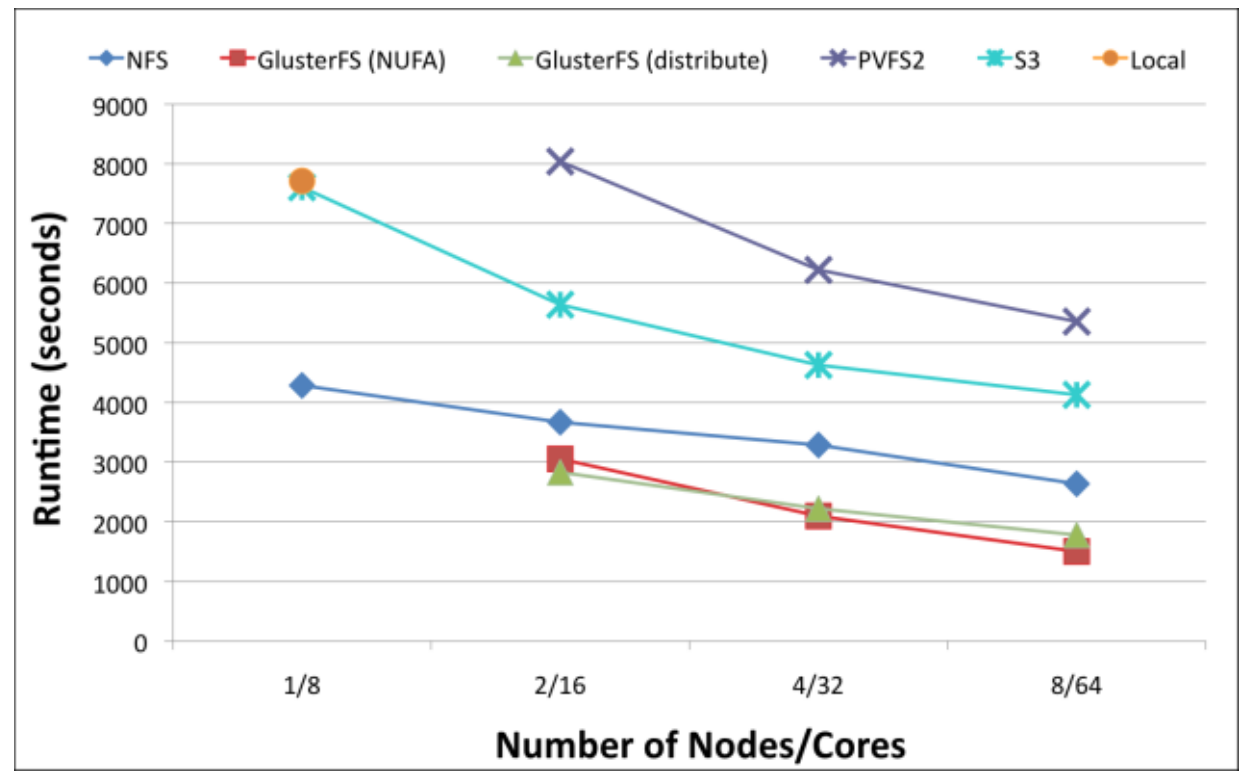

Fig. 2. Performance of Montage using different storage systems.

### B. Epigenome

The performance results for Epigenome are shown in Fig. 3. Epigenome is mostly CPU-bound, and performs relatively little I/O compared to Montage and Broadband. As a result, the choice of storage system has less of an impact on the performance of Epigenome compared to the other applications. In general, the performance was almost the same for all storage systems, with S3 and PVFS performing slightly worse than NFS and GlusterFS. Unlike Montage, for which NFS performed better than the local disk in the single node case, for Epigenome the local disk was significantly faster.

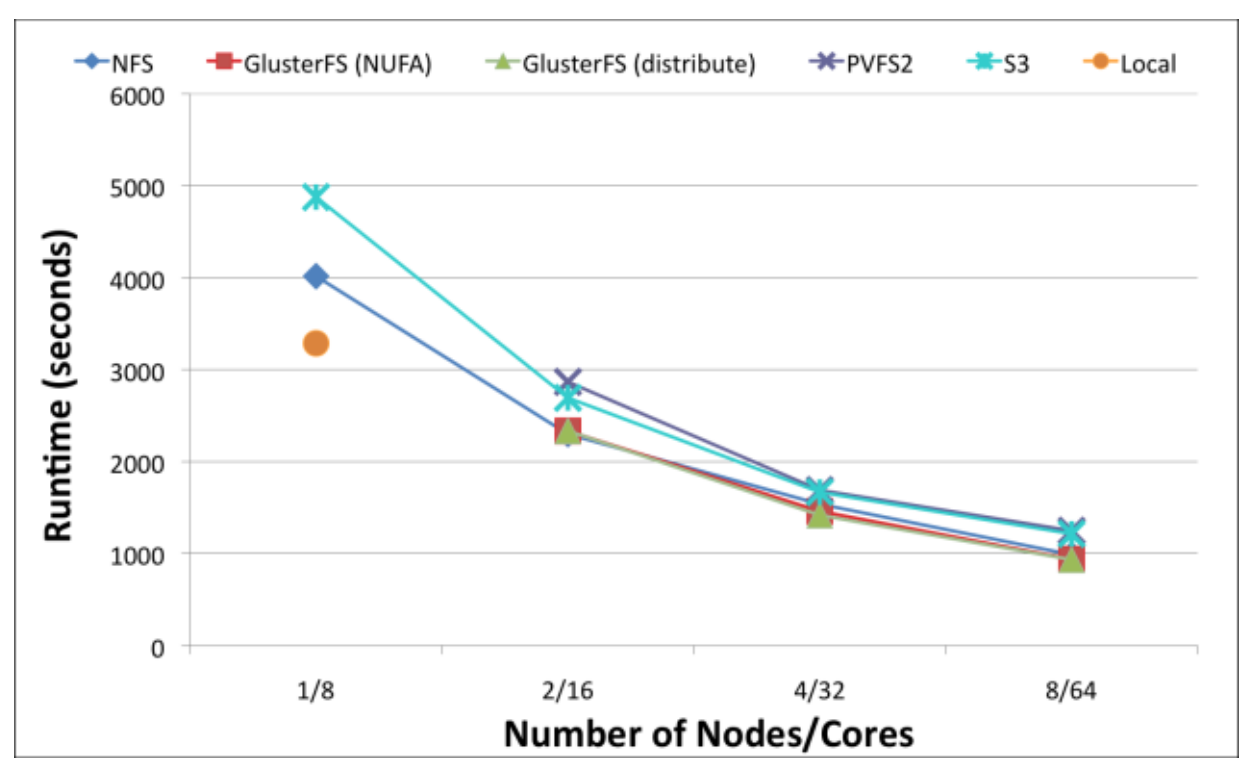

Fig. 3. Performance of Epigenome using different storage systems.

### C. Broadband

The performance results for Broadband are shown in Fig. 4. In contrast to the other applications, the best overall performance for Broadband was achieved using Amazon S3 and not GlusterFS. This is likely due to the fact that Broadband reuses many input files, which improves the effectiveness of the S3 client cache. Many of the transformations in Broadband consist of several executables that are run in sequence like a mini workflow. This would explain why GlusterFS (NUFA) results in better performance than GlusterFS (distribute). In the NUFA case all the outputs of a transformation are stored on the local disk, which results in much better locality for Broadband's workflow-like transformations. An additional Broadband experiment was run using a different NFS server (m2.4xlarge, 64 GB memory, 8 cores) to see if a more powerful server would significantly improve NFS performance. The result was better than the smaller server for the 4-node case (4368 seconds vs. 5363 seconds), but was still significantly worse than GlusterFS and S3 (<3000 seconds in all cases). The decrease in performance using NFS between 2 and 4 nodes was consistent across repeated experiments and was not affected by any of the NFS parameter changes we tried. Similar to Montage, Broadband appears to have relatively poor performance on PVFS, possibly because of the large number of small files it generates (>5,000).

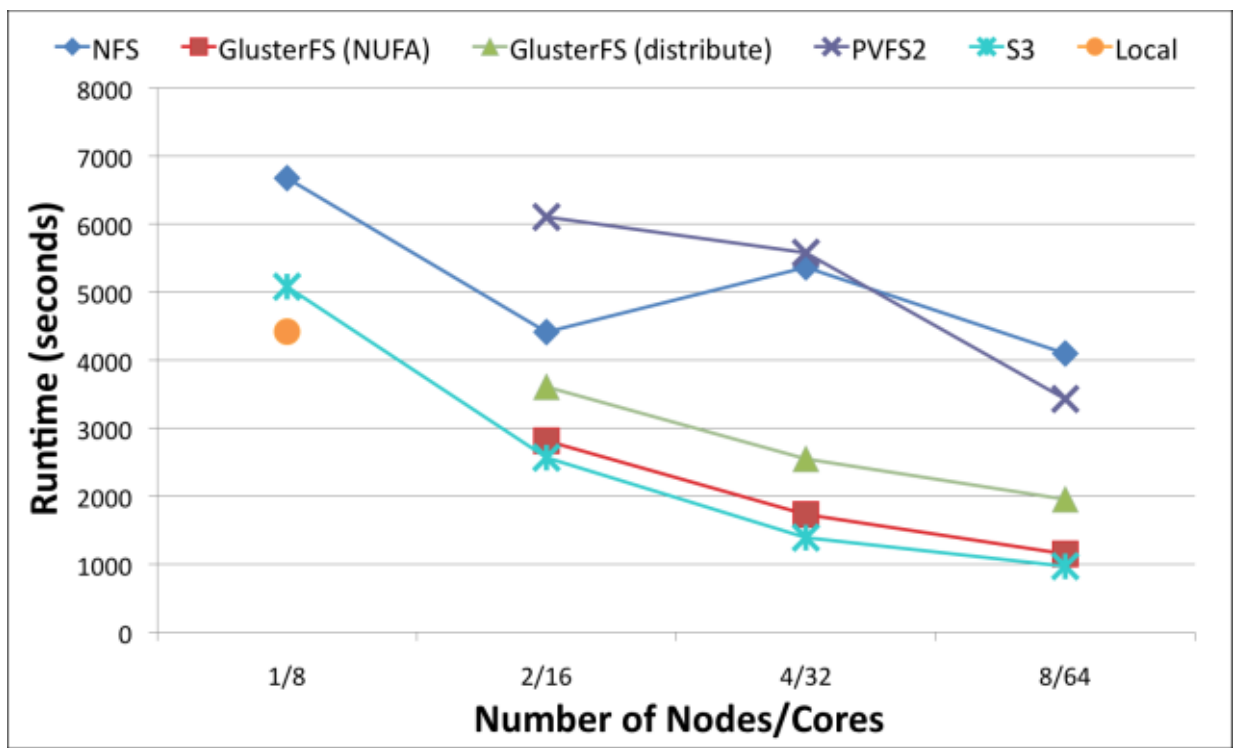

Fig. 4. Performance of Broadband using different storage systems.

## VI. Cost Comparison

In this section we analyze the cost of running workflow applications using the selected storage systems. There are three different cost categories when running an application on EC2. These include: resource cost, storage cost, and transfer cost. Resource cost includes charges for the use of VM instances in EC2; storage cost includes charges for keeping VM images and input data in S3 or EBS; and transfer cost includes charges for moving input data, output data and log files between the submit host and EC2. In our previous work in [16] we analyzed the storage and transfer costs for Montage, Broadband and Epigenome, as well as the resource cost on single nodes. In this paper we extend that analysis to multiple nodes based on our experiments with shared storage systems.

One important issue to consider when evaluating the cost of a workflow is the granularity at which the provider charges for resources. In the case of EC2, Amazon charges for resources by the hour, and any partial hours are rounded up. One important result of this is that there is no cost benefit to adding resources for workflows that run for less than an hour, even though doing so may improve runtime. Another result of this is that it is difficult to compare the costs of different solutions. In order to better illustrate the costs of the various storage systems we use two different ways to calculate the total cost of a workflow: per hour charges, and per second charges. Per hour charges are what Amazon actually charges for the usage, including rounding up to the nearest hour, and per second charges are what the experiments would cost if Amazon charged per second. We compute per second rates by dividing the hourly rate by 3,600 seconds.

It should be noted that the storage systems do not have the same cost profiles. NFS is at a disadvantage in terms of cost because of the extra node that was used to host the file system. This results in an extra cost of $0.68 per workflow for all applications. An alternative NFS configuration would be to overload one of the compute nodes to host the file system. However, in such a configuration the performance is likely to decrease, which may offset any cost savings. In addition, reducing the dedicated-node NFS cost by $0.68 still does not make it cheaper to use than the other systems. S3 is also at a disadvantage compared to the other systems because Amazon charges a fee to store data in S3. This fee is $0.01 per 1,000 PUT operations, $0.01 per 10,000 GET operations, and $0.15 per GB-month of storage (transfers are free within EC2). For Montage this results in an extra cost of $0.28, for Epigenome the extra cost is $0.01, and for Broadband the extra cost is $0.02. Note that the S3 cost is somewhat reduced by caching in the S3 client, and that the storage cost is insignificant for the applications tested (<< $0.01).

The total cost for Montage, Epigenome and Broadband, using both per-hour and per-second charges, and including extra charges for NFS and S3, is shown in Figs. 5-7. In addition to the cost of running on the storage systems described in Section IV, we also include the cost of running on a single node using the local disk (Local in the figures). For Montage the lowest cost solution was GlusterFS on two nodes. This is consistent with GlusterFS producing the best performance for Montage. For Epigenome the lowest cost solution was a single node using the local disk. Also notice that, because Epigenome is not I/O intensive, the difference in cost between the various storage solutions is relatively small. For Broadband the local disk, GlusterFS and S3 all tied for the lowest cost. For all of the applications the per-second cost was less than the per-hour cost—sometimes significantly less. This suggests that a cost-effective strategy would be to provision a virtual cluster and use it to run many workflows, rather than provisioning a virtual cluster for each workflow.

One final point to make about the cost of these experiments is the effect of adding resources. Assuming that resources have uniform cost and performance, in order for the cost of a workflow to decrease when resources are added the speedup of the application must be super-linear. Since this is rarely the case in any parallel application it is unlikely that there will ever be a cost benefit for adding resources, even though there may still be a performance benefit. In our experiments adding resources reduced the cost of a workflow for a given storage system in only 2 cases: 1 node to 2 nodes using NFS for both Epigenome and Broadband. In both of those cases the improvement was a result of the non-uniform cost of resources due to the extra node that was used for NFS. In all other cases the cost of the workflows only increased when resources were added. Assuming that cost is the only consideration and that resources are uniform, the best strategy is to either provision only one node for a workflow, or to use the fewest number of resources possible to achieve the required performance.

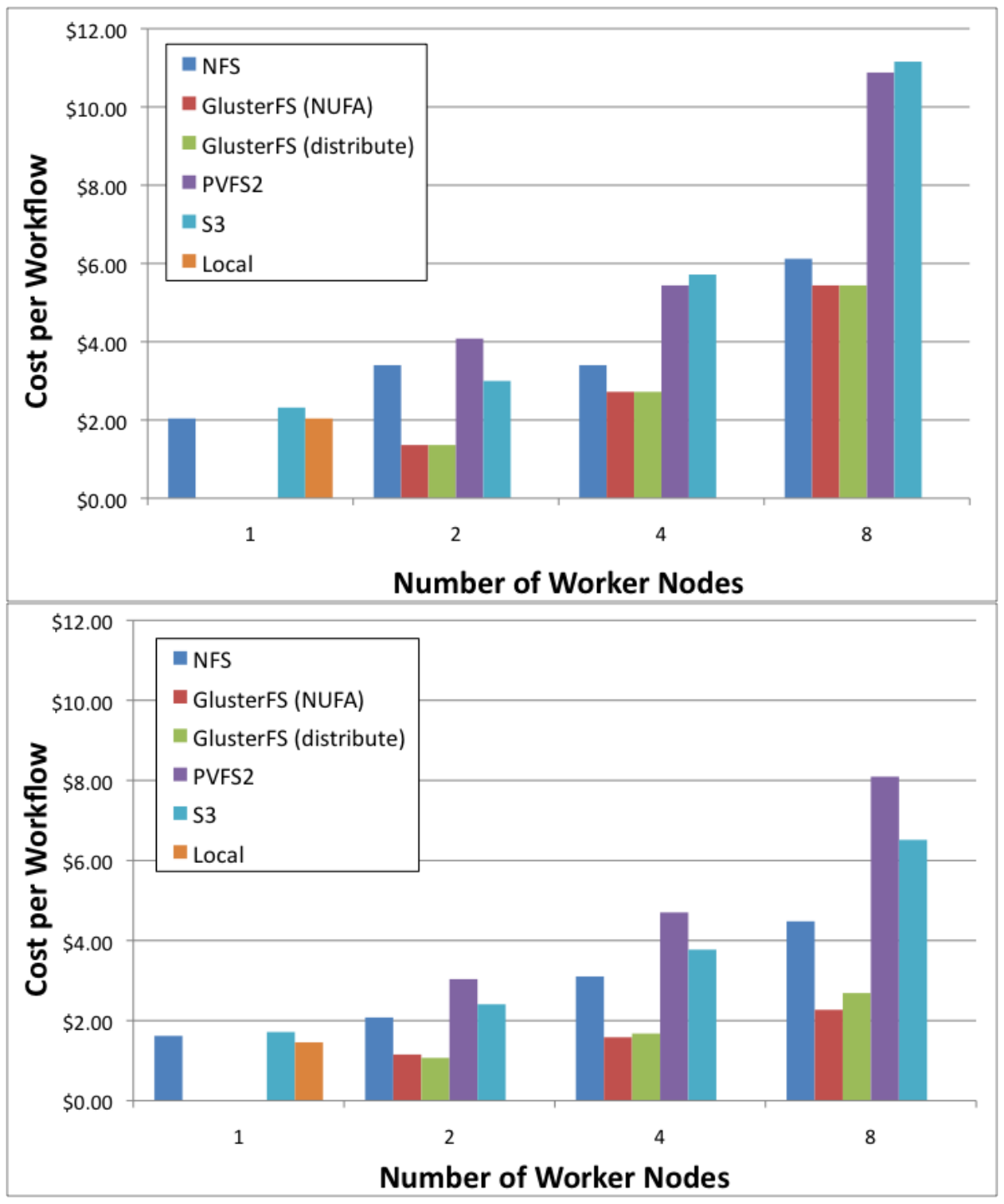

Fig. 5. Montage cost assuming per-hour charges (top) and per-second charges (bottom)

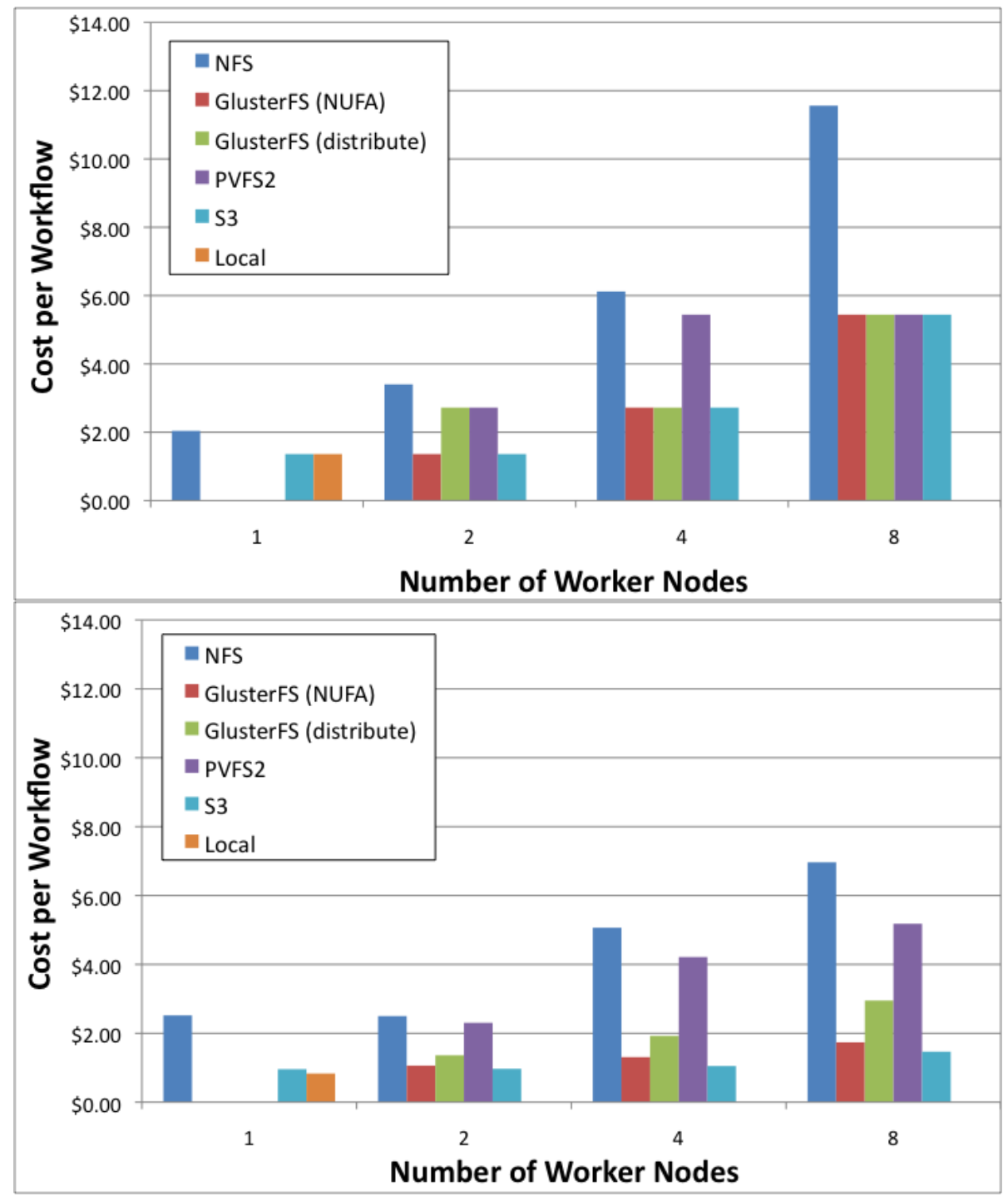

Fig. 7. Broadband cost assuming per-hour charges (top) and per-second charges (bottom)

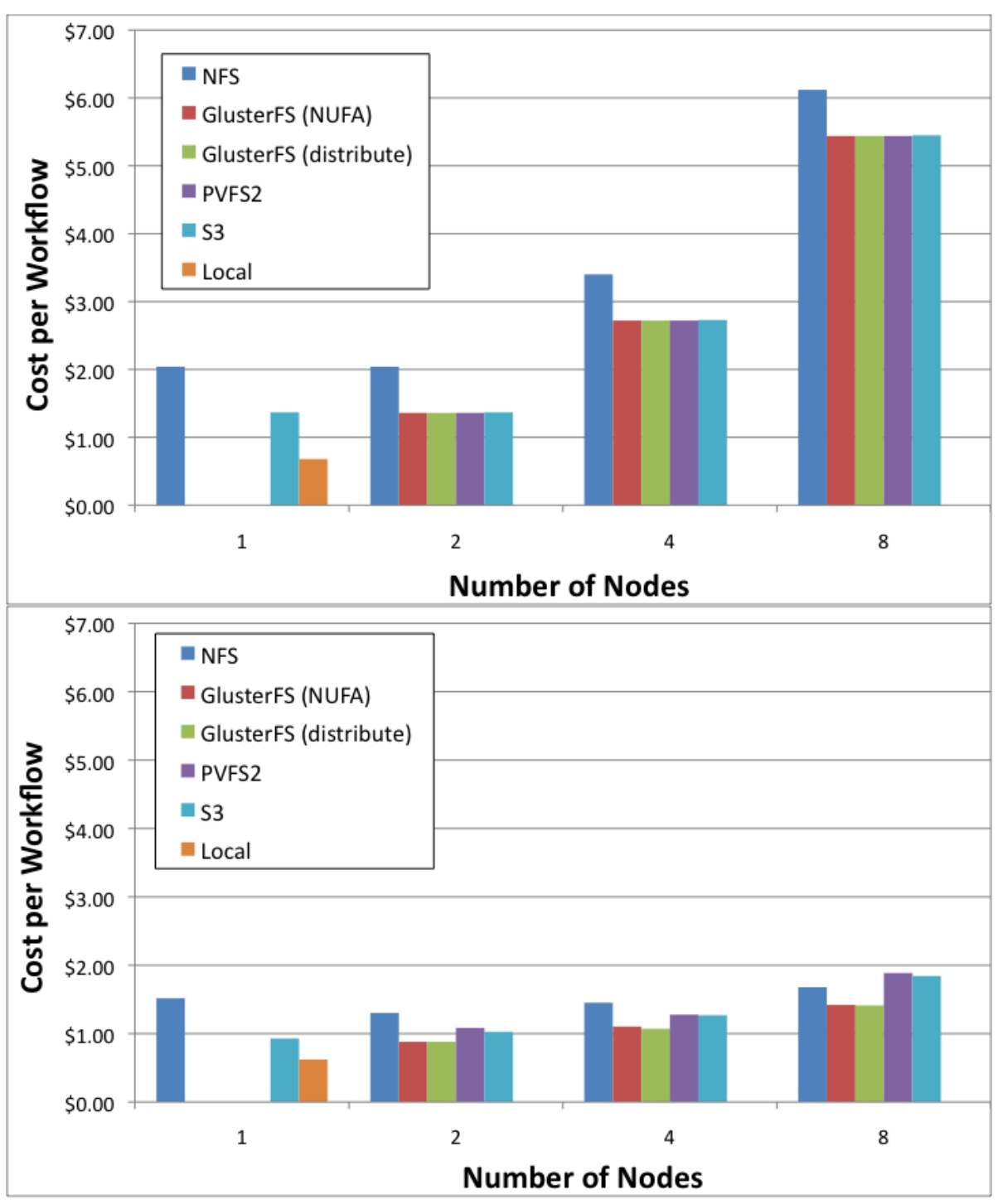

Fig. 6. Epigenome cost assuming per-hour charges (top) and per-second charges (bottom)

## VII. Related Work

Much previous research has investigated the performance of parallel scientific applications on virtualized and cloud platforms [8][9][13][22][25][26][32][34][35]. Our work differs from these in two ways. First, most of the previous efforts have focused on tightly-coupled applications such as MPI applications. In comparison, we have focused on scientific workflows, which are loosely-coupled parallel applications with very different requirements (although it is possible for individual workflow tasks to use MPI, we did not consider workflows with MPI tasks here). Second, previous efforts have focused mainly on micro benchmarks and benchmark suites such as the NAS parallel benchmarks [23]. Our work, on the other hand, has focused on the performance and cost of real-world applications.

Vecchiola, et al. have conducted research similar to our work [31]. They ran an fMRI workflow on Amazon EC2 using S3 for storage, compared the performance to Grid'5000, and analyzed the cost on different numbers of nodes. In comparison, our work is broader in scope. We use several applications from different domains with different resource requirements, and we experiment with five different storage systems.

In our own previous work on the use of cloud computing for workflows we have studied the cost and performance of clouds via simulation [6], using an experimental cloud [12], and using single EC2 nodes [16]. In this paper we have extended that work to consider larger numbers of resources and a variety of storage systems.

## VIII. CONCLUSION

In this paper we examined the performance and cost of several different storage systems that can be used to communicate data within a scientific workflow running on the cloud. We evaluated the performance and cost of three workflow applications representing diverse application domains and resource requirements on Amazon's EC2 platform using different numbers of resources (1-8 nodes corresponding to 8-64 cores) and five different storage systems. Overall we found that cloud platforms like EC2 do provide a good platform for deploying workflow applications.

One of the major factors inhibiting storage performance on EC2 is the first write penalty on ephemeral disks. We found that this significantly reduces the performance of storage systems deployed in EC2. This penalty seems to be unique to this execution platform. Repeating these experiments on another cloud platform may produce better results.

We found that the choice of storage system has a significant impact on workflow runtime. In general, GlusterFS delivered good performance for all the applications tested and seemed to perform well with both a large number of small files, and a large number of clients. S3 produced good performance for one application, possibly due to the use of caching in our implementation of the S3 client. NFS performed surprisingly well in cases where there were either few clients, or when the I/O requirements of the application were low. Both PVFS and S3 performed poorly on workflows with a large number of small files, although the version of PVFS we used did not contain optimizations for small files that were included in subsequent releases.

As expected, we found that cost closely follows performance. In general the storage systems that produced the best workflow runtimes resulted in the lowest cost. NFS was at a disadvantage compared to the other systems when it used an extra, dedicated node to host the file system, however, overloading a compute node would not have significantly reduced the cost. Similarly, S3 is at a disadvantage, especially for workflows with many files, because Amazon charges a fee per S3 transaction. For two of the applications (Montage, I/O-intensive; Epigenome CPU-intensive) the lowest cost was achieved with GlusterFS, and for the other application (Broadband—Memory-intensive) the lowest cost was achieved with S3.

Although the runtime of the applications tested improved when resources were added, the cost did not. This is a result of the fact that adding resources only improves cost if speedup is superlinear. Since that is rarely ever the case, it is better from a cost perspective to either provision one node to execute an application, or to provision the minimum number of nodes that will provide the desired performance. Also, since Amazon bills by the hour, it is more cost-effective to run for long-periods in order to amortize the cost of unused capacity. One way to achieve this is to provision a single virtual cluster and use it to run multiple workflows in succession.

In this work we only considered workflow environments in which a shared storage system was used to communicate data between workflow tasks. In the future we plan to investigate configurations in which files can be transferred directly from one computational node to another.

## ACKNOWLEDGEMENTS

This work was supported by the National Science Foundation under the SciFlow (CCF-0725332) and Pegasus (OCI-0722019) grants. This research made use of Montage, funded by the National Aeronautics and Space Administration's Earth Science Technology Office, Computation Technologies Project, under Cooperative Agreement Number NCC5-626 between NASA and the California Institute of Technology.

## REFERENCES

[1] Amazon.com, "Elastic Compute Cloud (EC2)," http://aws.amazon.com/ec2.

[2] Amazon.com, "Simple Storage Service (S3)," http://aws.amazon.com/s3.

[3] P. Carns, W. Ligon, R. Ross, and R. Thakur, "PVFS: A Parallel File System for Linux Clusters," *4th Annual Linux Showcase and Conference*, 2000, pp. 317-327.

[4] J.S. Chase, D.E. Irwin, L.E. Grit, J.D. Moore, and S.E. Sprenkle, "Dynamic virtual clusters in a grid site manager," 2003, pp. 90-100.

[5] "DAGMan: Directed acyclic graph manager," http://cs.wisc.edu/condor/dagman.

[6] E. Deelman, G. Singh, M. Livny, B. Berriman, and J. Good, "The Cost of Doing Science on the Cloud: The Montage Example," 2008.

[7] E. Deelman, G. Singh, M. Su, J. Blythe, Y. Gil, C. Kesselman, G. Mehta, K. Vahi, G.B. Berriman, J. Good, A. Laity, J.C. Jacob, and D.S. Katz, "Pegasus: A framework for mapping complex scientific workflows onto distributed systems," *Scientific Programming*, vol. 13, 2005, pp. 219-237.

[8] C. Evangelinos and C.N. Hill, "Cloud Computing for Parallel Scientific HPC Applications: Feasibility of Running Coupled Atmosphere-Ocean Climate Models on Amazon's EC2," 2008.

[9] R.J. Figueiredo, P.A. Dinda, and J.A. Fortes, "A case for grid computing on virtual machines," 2003, pp. 550-559.

[10] I. Foster, T. Freeman, K. Keahey, D. Scheftner, B. Sotomayer, and X. Zhang, "Virtual Clusters for Grid Communities," 2006, pp. 513-520.

[11] Gluster, Inc., "GlusterFS," http://www.gluster.org.

[12] C. Hoffa, G. Mehta, T. Freeman, E. Deelman, K. Keahey, B. Berriman, and J. Good, "On the Use of Cloud Computing for Scientific Workflows," *3rd International Workshop on Scientific Workflows and Business Workflow Standards in e-Science (SWBES '08)*, 2008.

[13] W. Huang, J. Liu, B. Abali, and D.K. Panda, "A Case for High Performance Computing with Virtual Machines," 2006.

[14] F. Hupfeld, T. Cortes, B. Kolbeck, J. Stender, E. Focht, M. Hess, J. Malo, J. Marti, and E. Cesario, "The XtreemFS architecture - a case for object-based file systems in Grids," *Concurrency and Computation: Practice & Experience*, vol. 20, 2008, pp. 2049-2060.

[15] Hyperic, Inc., "CloudStatus," http://www.cloudstatus.com.

[16] G. Juve, E. Deelman, K. Vahi, and G. Mehta, "Scientific Workflow Applications on Amazon EC2," *Workshop on Cloud-based Services and Applications in conjunction with 5th IEEE International Conference on e-Science (e-Science 2009)*, Oxford, UK: 2009.

[17] D.S. Katz, J.C. Jacob, E. Deelman, C. Kesselman, S. Gurmeet, S. Mei-Hui, G.B. Berriman, J. Good, A.C. Laity, and T.A. Prince, "A comparison of two methods for building astronomical image mosaics on a grid," *Proceedings of the*

*2005 International Conference on Parallel Processing Workshops (ICPPW 05)*, 2005, pp. 85-94.
[18] K. Keahey and T. Freeman, "Contextualization: Providing One-Click Virtual Clusters," *4th International Conference on eScience (eScience '08)*, 2008.
[19] H. Li, J. Ruan, and R. Durbin, "Mapping short DNA sequencing reads and calling variants using mapping quality scores," *Genome Research*, vol. 18, 2008, pp. 1851-1858.
[20] "Linux Software RAID," https://raid.wiki.kernel.org/index.php/Linux_Raid.
[21] M.J. Litzkow, M. Livny, and M.W. Mutka, "Condor: A Hunter of Idle Workstations," 1988, pp. 104-111.
[22] J. Napper and P. Bientinesi, "Can Cloud Computing Reach the Top500?," 2009.
[23] NASA Advanced Supercomputing Division, "NAS Parallel Benchmarks," http://www.nas.nasa.gov/Resources/Software/npb.html.
[24] Oracle Corporation, "Lustre parallel filesystem," http://www.lustre.org.
[25] S. Ostermann, A. Iosup, N. Yigitbasi, R. Prodan, T. Fahringer, and D. Epema, "A Performance Analysis of EC2 Cloud Computing Services for Scientific Computing," *Proceedings of Cloudcomp 2009*, Munich, Germany: 2009.
[26] M.R. Palankar, A. Iamnitchi, M. Ripeanu, and S. Garfinkel, "Amazon S3 for science grids: a viable solution?," 2008.
[27] "ptrace - process trace," *Linux Programmer's Manual*.
[28] R. Sandberg, D. Golgberg, S. Kleiman, D. Walsh, and B. Lyon, "Design and Implementation of the Sun Network Filesystem," *USENIX Conference Proceedings*, Berkeley, CA: 1985.
[29] Southern California Earthquake Center, "Community Modeling Environment (CME)," http://www.scec.org/cme.
[30] "USC Epigenome Center," http://epigenome.usc.edu.
[31] C. Vecchiola, S. Pandey, and R. Buyya, "High-Performance Cloud Computing: A View of Scientific Applications," *International Symposium on Parallel Architectures, Algorithms, and Networks*, 2009.
[32] E. Walker, "Benchmarking Amazon EC2 for High-Performance Scientific Computing," *Login*, vol. 33, pp. 18-23.
[33] S.A. Weil, S.A. Brandt, E.L. Miller, D.D.E. Long, and C. Maltzahn, "Ceph: A scalable, high-performance distributed file system," *7th Symposium on Operating Systems Design and Implementation (OSDI 06)*, 2006, pp. 307-320.
[34] L. Youseff, R. Wolski, B. Gorda, and C. Krintz, "Paravirtualization for HPC Systems," *Workshop on Xen in High-Performance Cluster and Grid Computing*, 2006.
[35] W. Yu and J. Vetter, "Xen-Based HPC: A Parallel I/O Perspective," *8th IEEE International Symposium on Cluster Computing and the Grid (CCGrid '08)*, 2008.